\documentclass[12pt, preprint]{aastex}
\usepackage[dvips]{color}
\newcommand{\re}{}
\newcommand{\bk}{\textcolor[rgb]{0,0,0}}

\shorttitle{Interstellar chloronium}
\shortauthors{Neufeld et al.}

\begin{document}

\title{Herschel$^*$ observations of interstellar chloronium}
\author{David A.~Neufeld\altaffilmark{1}, Evelyne Roueff\altaffilmark{2}, Ronald L.\ Snell\altaffilmark{3}, Dariusz Lis\altaffilmark{4}, Arnold O.\ Benz\altaffilmark{5}, Simon Bruderer\altaffilmark{6}, John H.\ Black\altaffilmark{7}, Massimo De Luca\altaffilmark{8}, Maryvonne Gerin\altaffilmark{8}, Paul F.\ Goldsmith\altaffilmark{9}, Harshal Gupta\altaffilmark{9}, Nick Indriolo\altaffilmark{1}, Jacques Le Bourlot\altaffilmark{2}, Franck Le Petit\altaffilmark{2}, Bengt Larsson\altaffilmark{7},  Gary J.\ Melnick\altaffilmark{10}, Karl M.\ Menten\altaffilmark{11}, Raquel Monje\altaffilmark{4}, Zs\'ofia Nagy\altaffilmark{12}, Thomas G.\ Phillips\altaffilmark{4}, Aage Sandqvist\altaffilmark{13}, Paule Sonnentrucker\altaffilmark{14}, Floris van der Tak\altaffilmark{12,15}, and Mark~G.~Wolfire\altaffilmark{16}}

\altaffiltext{*}{Herschel is an ESA space observatory with science instruments provided
by European-led Principal Investigator consortia and with important
participation from NASA}
\altaffiltext{1}{Dept.\ of Physics \& Astronomy, Johns Hopkins Univ.\,
3400~N.~Charles~St., Baltimore, MD 21218, USA}
\altaffiltext{2}{Observatoire de Paris-Meudon, LUTH UMR 8102, 5 Pl.\ Jules Janssen,
F-92195 Meudon Cedex, France.}
\altaffiltext{3}{Astronomy Department, University of Massachusetts at Amherst, Amherst, MA 01003, USA}
\altaffiltext{4}{California Institute of Technology, 1200 East California Boulevard, Pasadena, CA 91125, USA} 
\altaffiltext{5}{Institute of Astronomy, ETH Zurich, 8092 Zurich, Switzerland}
\altaffiltext{6}{Max Planck Institut für Extraterrestrische Physik, Giessenbachstrasse 1, 85748, Garching, Germany}
\altaffiltext{7}{Dept.\ of Earth and Space Sciences, Chalmers University of Technology, Onsala, Sweden}
\altaffiltext{8}{LERMA, UMR 8112 du CNRS, Observatoire de Paris, \'Ecole Normale Sup\'erieure, UPMC \& UCP, France}
\altaffiltext{9}{JPL, California Institute of Technology, Pasadena, USA}
\altaffiltext{10}{Harvard-Smithsonian Center for Astrophysics, 60 Garden St.\ , Cambridge, MA 02138}
\altaffiltext{11}{Max-Planck-Institut f\"ur Radioastronomie, Auf dem H\"ugel 69, 53121 Bonn, Germany}
\altaffiltext{12}{University of Groningen, Groningen, The Netherlands}
\altaffiltext{13}{Stockholm Observatory, Stockholm University, AlbaNova University Center, Stockholm, Sweden}
\altaffiltext{14}{Space Telescope Science Institute, 3700 San Martin Drive, Baltimore, MD 21218, USA}
\altaffiltext{15}{SRON Netherlands Institute for Space Research}
\altaffiltext{16}{Department of Astronomy, University of Maryland, College Park, MD 20742, USA}

\begin{abstract}

{\re Using the {\it Herschel Space Observatory}'s Heterodyne Instrument for the Far-Infrared (HIFI)}, we have observed para-chloronium (H$_2$Cl$^+$) toward six sources in the Galaxy.  
%These observations were performed, using the Band 1a receiver in dual beam switch mode, as part of the PRISMAS  (``Probing Interstellar Molecules with Absorption Line Studies'') and HOP(``Herschel Oxygen Project'') Key Programs.  
We detected interstellar chloronium absorption in foreground molecular clouds along the sight-lines to the bright submillimeter continuum sources Sgr A (+50 km/s cloud) and W31C.  Both the para-H$_2^{35}$Cl$^+$ and para-H$_2^{37}$Cl$^+$ isotopologues were detected, {\re through observations of their $1_{11}-0_{00}$ transitions at rest frequencies of 485.42 and 484.23~GHz, respectively.}  For an assumed ortho-to-para ratio of 3, the observed optical depths imply that chloronium accounts for $\sim 4 - 12 \%$ of chlorine nuclei in the gas phase.  We detected interstellar chloronium emission from two sources in the Orion Molecular Cloud 1: the Orion Bar photodissociation region and the Orion South condensation.  For an assumed ortho-to-para ratio of 3 {\re for chloronium}, the observed emission line fluxes imply total beam-averaged column densities of $\sim 2 \times 10^{13} \rm \, cm^{-2}$ and $\sim 1.2 \times 10^{13} \rm \, cm^{-2}$, respectively, for chloronium in these two sources.  We obtained upper limits on the para-H$_2^{35}$Cl$^+$ line strengths toward H$_2$ Peak 1 in the Orion Molecular cloud and toward the massive young star AFGL 2591.
The chloronium abundances inferred in this study are typically at least a factor $\sim 10$ larger than the predictions of steady-state theoretical models for the chemistry of interstellar molecules containing chlorine.  {\re Several explanations for this discrepancy were investigated, but none has proven satisfactory, and thus the large observed abundances of chloronium remain puzzling.}

\end{abstract}

\keywords{ISM:~molecules -- Submillimeter:~ISM -- Molecular processes }

\section{Introduction}

With solar abundances estimated as $3.6 \times 10^{-8}$ and $3.2 \times 10^{-7}$ respectively relative to hydrogen (Asplund et al.\ 2009), fluorine and chlorine are only minor constituents of the Universe, with abundances 3 -- 4 orders of magnitude smaller than those of carbon, nitrogen and oxygen.  Nevertheless, these halogen elements show a strong tendency toward the formation of hydrides within the interstellar medium, with the surprising result that hydrogen fluoride (HF) and chloronium (H$_2$Cl$^+$) can exhibit abundances that are comparable to those of CH and H$_2$O within diffuse molecular clouds (Neufeld et al.\ 2010a, Sonnentrucker et al.\ 2010, Lis et al.\ 2010).  This tendency is a consequence of a unique characteristic of the thermochemistry of these elements: they are the only elements that can react exothermically with H$_2$ when in their predominant ionization stage in atomic clouds (i.e. atomic F, in the case of fluorine, and Cl$^+$, in the case of chlorine).  These reactions lead respectively to HF, which becomes the dominant reservoir of fluorine in diffuse molecular clouds (Neufeld et al.\ 2010; Sonnentrucker et al.\ 2010), and HCl$^+$, the tentative detection of which has been reported recently  (De Luca et al.\ 2011).  

HF is relatively inert, being destroyed slowly via photodissociation and reaction with C$^+$.
HCl$^+$, by contrast, can be destroyed by dissociative recombination, or by exothermic reaction with H$_2$.  The latter process leads to interstellar chloronium, discovered recently by Lis et al.\ (2010), who used the {\em Herschel 
Space Observatory} to perform absorption line spectroscopy of the $1_{11}-0_{00}$ transition of para-chloronium toward the bright submillimeter continuum sources NGC 6634I and Sgr B2 (S).  Chloronium is an intermediary in the production of hydrogen chloride (Neufeld \& Wolfire 2009, and references therein), the first chlorine-bearing {\re molecule} to be detected in the interstellar medium (Blake, Keene \& Phillips 1985); the latter is formed as a 
result of dissociative recombination of H$_2$Cl$^+$, or -- in dense clouds -- following proton transfer from H$_2$Cl$^+$ to CO. 

In this paper, we report four additional detections of chloronium obtained with {\em Herschel}'s Heterodyne Instrument for the Far-Infrared (HIFI; de Graauw et al.\ 2010) in two separate Key Programs.  We have observed the $1_{11}-0_{00}$ transition of para-H$_2$Cl$^+$, which lies less than 2 GHz from a molecular oxygen line that has been targeted in very deep integrations towards many sources in the HOP (``Herschel Oxygen Project''; {\re Co-PI's, P.\ Goldsmith \& R.\ Liseau}) Open Time Key Program.  Although studying H$_2$Cl$^+$ was not the primary goal of HOP, we ensured that chloronium was observed simultaneously with O$_2$ by a suitable choice of local oscillator setting, thus performing sensitive searches for chloronium towards all HOP sources.  The HOP source list includes a wide diversity of molecular environments, and the new detections of H$_2$Cl$^+$ reported here were obtained toward (1) the Orion Bar, a photodissociation region where intense radiation from the Trapezium stars in Orion is incident upon a dense molecular cloud; (2) Orion South, a dense condensation within the Orion Molecular ridge; and (3) the so-called ``+50 km/s cloud'' in the vicinity of Sgr A, a source of {\re far-infrared continuum radiation -- abutting the Sgr A East supernova remnant --} with a sight-line that is intersected by multiple foreground molecular clouds located between the Sun and the Galactic Center.  In addition, the $1_{11}-0_{00}$ transition of para-H$_2$Cl$^+$ was targeted in observations of the massive star-forming region W31C, performed as part of the PRISMAS  (``Probing Interstellar Molecules with Absorption Line Studies''; P.I., M. Gerin) Guaranteed Time Key Program.  W31C (also known as G10.6--0.4) is a bright continuum source, located at an  estimated distance of $4.8^{+0.4}_{-0.8}$~kpc (Fish et al.\ 2003) within the so-called ``--30~km/s'' spiral arm, and provides a sight-line that intersects several foreground molecular clouds, the arrangement of which has been elucidated by Corbel \& Eikenberry (2004; see their Figure 8).    Because W31C shows a large continuum flux at infrared and radio wavelengths, this sight-line has proven one of the most valuable in the Galaxy for the study of interstellar gas by absorption line spectroscopy.  In addition to the new detections of chloronium, upper limits on the $1_{11}-0_{00}$ transition have been obtained {\re from the HOP program} toward H$_2$ Peak 1 in Orion, a source of shock-excited molecular hydrogen emission, and toward the young massive star AFGL 2591, {\re proposed to be a strong far-ultraviolet emitter that irradiates outflow cavity walls within a dense infalling envelope (Bruderer et al.\ 2010).}

The observations and data reduction methods are described in \S 2 below.  In \S 3, we present the resultant spectra and report detections of H$_2$Cl$^+$ toward four sources.  In \S 4 and \S 5, the column densities and abundances inferred from these detections are discussed in the context of models for the chemistry of chlorine-containing interstellar molecules.

\section{Observations and data reduction}

All the observations reported here were obtained using the HIFI instrument in dual beam switch (DBS) mode, with the $1_{11}-0_{00}$ transition of para-chloronium in the lower sideband of the Band 1a receiver.  Table 1 lists the observations that led to detections\footnote{Upper limits were obtained toward Orion H$_2$ Peak 1 ($\alpha = \rm  5h\, 35m\, 13.70\, s, \delta = -5d \, 22^{\prime} \, 09.00^{\prime\prime}$ J2000) and AFGL 2591 ($\rm \alpha = 20h \, 29m \, 24.70\, s, \delta = +40d \, 11^{\prime} \, 19.00^{\prime\prime} J2000)$}.  The telescope beam was centered at the coordinates given in Table 1, with the reference positions located at offsets of 3$^\prime$ on either side of each source.  The beam size was 44$^{\prime\prime}$ (half power beam width).   The data were acquired using the Wide Band Spectrometer (WBS), which provides a spectral resolution of 1.1~MHz and a bandwidth of $\sim 4$~GHz.  For each observation, the date and total duration are given in Table 1, together with the r.m.s.\ noise achieved in a 1.1~MHz channel.
As discussed in Neufeld et al.\ (2010a), the data were processed using the standard HIFI pipeline to Level 2, providing fully calibrated spectra with the intensities expressed as antenna temperature and the frequencies in the frame of the Local Standard of Rest (LSR).
For the observations of W31C and Sgr A (+50 km/s cloud), we used three local oscillator (LO) frequencies, separated by a small offset, to confirm the assignment of any observed spectral feature to either the upper or lower sideband of the (double side band) HIFI receivers.  The resultant spectra were coadded so as to recover the signal-to-noise ratio that would have been obtained at a single LO setting.  Spectra obtained for the horizontal and vertical polarizations were found to be very similar in their appearance and noise characteristics and were likewise coadded.  In the case of the Orion Bar and Orion S, the data were acquired in the frequency scan observing mode, which automatically performs the observations at multiple LO settings.  Here,  sideband deconvolution was performed using the {\it doDeconvolution} tool in the Herschel Interactive Processing Environment (HIPE).

The relevant spectroscopy for chloronium is well established, thanks to direct laboratory measurements of its 180 -- 500 GHz rotational spectrum (Araki et al.\ 2001).  For the more abundant H$_2^{35}$Cl$^+$ isotopologue, the strongest ($F=5/2 - 3/2$) hyperfine component of the $1_{11} - 0_{00}$ transition has a measured frequency of $485417.670 \pm 0.015$~MHz, with the $F=3/2 - 3/2$ and $F=1/2 - 3/2$ components lying at frequency offsets of --4.24 and +3.13 MHz, respectively, corresponding to Doppler velocities of +2.6 and --1.9 km/s.  For H$_2^{37}$Cl$^+$, the $1_{11} - 0_{00}$ $F=5/2 - 3/2$ transition has a measured frequency of $484231.804 \pm 0.026$~MHz, and the $F=3/2 - 3/2$ and $F=1/2 - 3/2$ components lie at velocity offsets of of +2.1 and --1.7 km/s.  The dipole moment of chloronium has been estimated as 1.89 Debye in {\it ab initio} calculations performed by M\"uller (2008).

\section{Results}

Clear detections of the $\rm H_2^{35}Cl^+$ $1_{11}-0_{00}$ transition were obtained in absorption toward Sgr A (+50 km/s cloud) and W31C, and in emission toward the Orion Bar and Orion South.   In addition, the $1_{11}-0_{00}$ line of the $\rm H_2^{37}Cl^+$ isotopologue was detected toward Sgr A (+50 km/s cloud) and W31C, but this transition was outside the spectral region covered by the Orion Bar and Orion S observations.  

Figure 1 shows the spectra of H$_2^{35}$Cl$^+$ (and, for two sources, H$_2^{37}$Cl$^+$) obtained toward the four sources Sgr A (+50 km/s cloud), W31C, the Orion Bar, and Orion South.   Here, the intensity is plotted on the scale of antenna temperature, as a function of the LSR velocity for the strongest hyperfine component.  For the Orion Bar and Orion S, Figure 1 shows the continuum-subtracted sideband-deconvolved spectra, whilst for Sgr A (+50 km/s cloud) and W31C, Figure 1 shows double sideband (DSB) spectra without continuum subtraction.   In DSB spectra, the complete absorption of radiation in the observed spectral lines would reduce the antenna temperature to one-half its continuum value (given a sideband gain ratio of unity).  The {\it single sideband} continuum antenna temperatures are listed in Table 1.  {\re For the Orion Bar and Orion South, the frequency-integrated line fluxes are also listed in units of K $\rm km\,s^{-1}$, along with the line centroids and line widths.  For the latter two parameters, two fitting methods were used: in the first method, the spectrum was fit as the sum of three Gaussians, of identical width, with the centroids separated in accord with the hyperfine splitting for the H$_2^{35}$Cl$^+$ transition, and with relative strengths appropriate to L.T.E.; in the second method, a single-component Gaussian fit was adopted, computed on the velocity scale for the strongest hyperfine component.  The second method, of course, leads to line widths that are broader because of the effects of hyperfine splitting, and also results in a $\sim 0.8 \rm \,km\,s^{-1}$ shift in the centroid velocity.  In the Orion bar, the former method yields a line centroid ($10.5\, \rm km \, s^{-1}$) that is in excellent accord with that observed for other spectral lines, although the line width ($\sim 3.4\, \rm km \, s^{-1}$ FWHM) is somewhat larger than those typically observed for optically-thin molecules along this sight-line (e.g. Parise et al.\ 2009), arguing, perhaps, in favor of a surface origin for H$_2$Cl$^+$.  Toward Orion S, good agreement is obtained with the line centroid ($7.2\rm \,km\,s^{-1}$) and width ($3.7\,\rm km\,s^{-1}$) inferred by Rodr{\'{\i}}guez-Franco et al.\ (2001) from observations of CN $N = 3- 2$.}

Toward W31C, several interloper lines are present {\re in the H$_2^{37}$Cl$^+$ spectrum}.  In particular, the SO$_2 \,\, 13_{3,11}-12_{2,10}$, SO$_2 \,\,27_{1,27}-26_{0,26}$ and NS $^2\Pi_{1/2}\, J = 21/2 - 19/2$ emission lines appear respectively near --25 +2, and $+48 \, \rm km \, s^{-1}$.  The latter (NS 484.151~GHz) transition, in particular, affects our ability to determine the H$_2^{37}$Cl$^+$column density at LSR velocities above $\sim 40\,\rm km\,s^{-1}$. However, this NS transition is a member of a lambda doublet, with another component appearing at 484.547~GHz (not shown in Figure 1).  We have attempted to remove the interfering doublet member by subtracting a feature with the same strength and profile observed for the 484.547~GHz component; the result of this subtraction appears as the blue histogram in Figure 1, and has been used in our subsequent determination of the H$_2^{37}$Cl$^+$ column density.

Toward Orion H$_2$ Peak 1, the effective sensitivity was significantly compromised by the presence of several interfering spectral lines, the influence of which -- at this position -- is probably a consequence of the beam partially overlapping the nearby Orion hot core (a rich emission line source located at an offset of 27$^{\prime\prime}$ from the beam center.)  Based upon the typical residuals remaining after the removal of known transitions of dimethyl ether and $\rm ^{33}SO_2$, we obtained an upper limit of $\sim 0.6$~K~km/s for the integrated $\rm H_2^{35}Cl^+$ $1_{11}-0_{00}$ line intensity.  Toward AFGL 2951, there is no evidence for emission or absorption at any Doppler velocity in a spectrum with a continuum antenna temperature of 57~mK and an r.m.s. noise of 4 mK (in a 1 km/s channel).

\section{Inferred column densities {\re and abundances of chloronium}}

\subsection{Chloronium absorption observed toward W31C and Sgr A}

The Sgr A (+50 km/s cloud) and W31C spectra show H$_2$Cl$^+$ absorption over a wide range of velocities, a behavior that is typical of other molecules observed in absorption toward these sources.  The observed molecular absorption is believed to arise largely in foreground diffuse clouds, unassociated with the background sources of continuum radiation.  

In Figures 2 and 3, we present the observed flux for the two sources showing H$_2^{35}$Cl$^+$ and H$_2^{37}$Cl$^+$ in {\it absorption}, now normalized with respect to the continuum flux in a single sideband.  Here, we assume a sideband gain ratio of unity (Roelfsema et al.\ 2012), and thus we plot the quantity $2T_A/T_A({\rm cont}) - 1$.  Using the procedure described by Neufeld et al.\ (2010b), we have deconvolved the hyperfine structure to obtain the spectra shown in red; these are the spectra that would have resulted if the hyperfine splitting were zero.
For comparison, the spectra (see figure captions for the sources of these spectra) of four other molecules detected with HIFI are shown: CH (the 536.761 GHz transition in green),  HF (the 1232.476 GHz transition in brown), para-H$_2$O (the 1113.343 GHz transition in magenta), and OH$^+$ (the 971.804 GHz transition in blue).  In addition, the H~I  21~cm spectrum is shown in dark green.  Vertical offsets have been introduced for clarity.

The spectra shown in Figures 2 and 3 indicate that all the absorbing species show broadly similar distributions in velocity space, but with HF, H$_2$O, and CH -- all believed to be good tracers of molecular hydrogen -- showing the strongest {\re mutual} resemblance.  Toward the Sgr A (+50 km/s cloud), chloronium is apparently distinctive among the molecules in exhibiting a lack of absorption at LSR velocities smaller than $\sim - 70$~km/s ({\re and specifically a lack of absorption at $\sim - 130$~km/s, which normally arises within the Expanding Molecular Ring (EMR) in the inner region of the Galaxy.)
In this regard, the distribution of $\rm H_2Cl^+$ in velocity space is more similar to that of atomic hydrogen than that of any of the other molecules shown in Figure 3.  Although, given the sensitivity of the observations,  this behavior is only tentatively established, it argues in favor of theoretical predictions (\S see 5.1 below) that the chloronium abundance is largest within clouds that are primarily atomic.}

In Table 2, the column densities of para-H$_2^{35}$Cl$^+$ and para-H$_2^{37}$Cl$^+$ are given for various velocity ranges, computed with the approximation that all para-chloronium molecules are in the ground $0_{00}$ state.  As for other hydrides observed toward these sources, that approximation is justified by the large spontaneous radiative rate for the observed transition\footnote{The Einstein A-coefficient for the observed para-H$_2^{35}$Cl$^+$ transition is $1.59 \times 10^{-2}\, \rm s^{-1}$, for an assumed dipole moment of 1.89~D.  This implies that the H$_2$ and electron ``critical densities'' (at which the rate of collisional deexcitation is equal to this spontaneous radiative decay rate) are $\sim 10^8\,\rm cm^{-3}$ and  $\sim 10^4\,\rm cm^{-3}$ respectively, { given the estimated collisional rate coefficients discussed in \S4.2}}, the low density of the foreground clouds responsible for the absorption, and the weak continuum radiation field at the frequency of the observed transition and at the location of the absorbing material.  Table 2 also presents the total $\rm H_2Cl^+$ column density (both isotopologues) for an assumed ortho-to-para ratio (OPR) of 3.  For each velocity range, the H~I  column density is also given, determined from 21 cm absorption spectra (Fish et al.\ 2003 for W31C; Lang et al.\ 2010 for Sgr A +50 km/s cloud) for an assumed H~I  spin temperature of 100~K.  Since the observed H$_2$Cl$^+$ is expected to arise primarily in clouds of small H$_2$ fraction (see \S 5.1 below), we also give the $N({\rm H_2Cl^+})/N({\rm H})$ abundance ratios in Table 2.  The mean value obtained for the 0 to 50 km/s LSR velocity range in W31C, $N({\rm H_2Cl^+})/N({\rm H\,I}) = 1.15 \times 10^{-8}$, corresponds to $\sim 12 \%$ of the gas-phase chlorine abundance 
within diffuse atomic clouds ($1.0 \times 10^{-7}$; Moomey, Federman \& Sheffer 2011); that obtained for the --70 to --30 km/s LSR velocity range in Sgr A (+50 km/s cloud), $N({\rm H_2Cl^+})/N({\rm H\,I}) = 4.4 \times 10^{-9}$, corresponds to $\sim 4 \%$ of the gas-phase chlorine abundance.  {\re For most velocity intervals, in both sources, the H$_2^{35}$Cl$^+$/H$_2^{37}$Cl$^+$ abundance ratio is similar to the $^{35}$Cl$^+$/$^{37}$Cl$^+$ isotopic ratio of 3 observed in the solar system (Lodders 2003).}

\subsection{Chloronium emission observed toward the Orion Bar}

The absorption line observations discussed above provide robust estimates of the H$_2$Cl$^+$ column densities, because almost all para-chloronium molecules are in the ground $0_{00}$ state (see \S 4.1 above), but the interpretation of the {\it emission line} observations is less straightforward.  At the densities of the H$_2$Cl$^+$ emission region in the Orion Bar -- $n({\rm H}_2) \sim 3 \times 10^4\,\rm cm^{-3}$ and $n({\rm e}) \sim 5 \,\rm cm^{-3}$ (discussed further below) -- the molecule will be subthermally-excited, with most para-H$_2$Cl$^+$ molecules in the ground $0_{00}$ state, and thus the emission line intensity will depend crucially upon the rate coefficients for collisional excitation of H$_2$Cl$^+$.  Because all excitations, to any excited rotational state, will be followed by a radiative cascade resulting in the emission of a $1_{11}-0_{00}$ line photon, the relevant parameter is the total rate of collisional excitation out of the ground rotational state.  As far as we are aware, no detailed calculations of the latter have been performed.  For excitation by H$_2$, we therefore adopt the results obtained (Faure et al.\ 2007) for the isovalent para-H$_2$O molecule, which yield a total rate coefficient for excitation from $0_{00}$ that is well-approximated by $q_{\rm tot}({\rm H}_2) = 2 \times 10^{-10} T_2^{0.5} \exp [-E(1_{11})/kT] \, \rm cm^3\ s^{-1}$, where $E(1_{11})$ is the energy of the lowest excited state ($1_{11}$), $T$ is the kinetic temperature, and $T_2 = T/100$~K.
%For excitation by helium, results of (REF) for para-H$_2$O yield an analogous expression: $q_{\rm tot}({\rm He}) = 0 \times 10^{-10} T_2^{0.0} \exp (-E[1_{11}]/kT)$  
For the case of collisional excitation by electrons, we used the Coulomb-Born approximation to relate the spontaneous radiative rate for the $1_{11}-0_{00}$ transition to the rate of electron impact excitation.  With the aid of equations 15, 12, and 11 in Neufeld \& Dalgarno (1989), we thereby obtained an estimate for the rate coefficient for excitation from $0_{00}$ that can be fit by $q_{\rm tot}({\rm e})= 2.6 \times 10^{-6} T_2^{-0.37} \exp [-E(1_{11})/kT] \, \rm cm^3\ s^{-1},$ where $E(1_{11})= 23$~K for the case of H$_2$Cl$^+$.   We have also considered -- and then rejected -- the possibility that excited states of H$_2$Cl$^+$ might be populated significantly when the molecule is formed.  Even if every H$_2$Cl$^+$ is formed in an excited state, this so-called ``formation pumping" can be shown to be negligible by means of an argument that equates the rate of formation of H$_2$Cl$^+$ to the rate of its destruction through dissociative recombination and proton transfer to neutral species.\footnote{\re The H$_2$Cl$^+$ destruction rate is $k_{\rm DR} n_e + k_{\rm PT} n({\rm X}),$ where $k_{\rm DR} = { 1.2} \times 10^{-7} (T/{\rm { 300} \, K})^{-0.85}\,\rm cm^3\, s^{-1}$ (Geppert \& Hamburg 2009, personal communication) is the rate coefficient for dissociative recombination of H$_2$Cl$^+$, $k_{PT} \sim 10^{-9}\rm \, cm^3\, s^{-1}$ is a typical rate coefficient for proton transfer, and $n({\rm X}) \sim {\rm few} \times 10^{-4} n({\rm H}_2)$ is the density of neutral species (e.g.\ CO) to which proton transfer might occur.  The first term in this expression is smaller than the rate of electron-impact excitation, whilst the second term is much smaller than the rate of collisional excitation by H$_2$ or H.}

The molecular hydrogen density in the Orion Bar has been estimated previously by means of PDR (Photodissociation Region) modeling (e.g.\ Jansen et al.\ 1995).  The consensus estimate, which we adopt here, implies a total density of hydrogen nuclei of $n_{\rm H}\sim 6 \times 10^4 \, \rm cm^{-3}$, although previous studies have suggested the presence of denser clumps with a small filling factor (e.g. Lis \& Schilke 2003); in the present work, we neglect the latter, because their contribution to the overall emission is expected to be small.  Theoretical studies for the chemistry of chlorine-bearing molecules (e.g.\ Neufeld \& Wolfire 2009, hereafter NW09) suggest that H$_2$Cl$^+$ is most abundant near the cloud surface where carbon is fully ionized, hydrogen is largely atomic, and the gas temperature is $\sim 500$~K.  In this region, the atomic hydrogen density is essentially the hydrogen nucleon density, $n({\rm H\,I}) \sim n_{\rm H}$, and  we estimate the collisional excitation rate (per H$_2$Cl$^+$ molecule) as $R_{\rm surface} = q_{\rm tot}({\rm H_2}) n_{\rm H} +  q_{\rm tot}({\rm e}) n_{\rm e} = {\re 3.8} \times 10^{-5}\rm \, s^{-1}$.  Here, we assumed that the rate of collisional excitation by H is equal to that for H$_2$, and we adopted an electron abundance equal to the gas-phase abundance of carbon ($\sim 1.6 \times 10^{-4}$ relative to H nuclei; Sofia et al.\ 2004).  We also considered the conditions existing far from the cloud surface, in the cloud interior, where the fractional ionization is low, hydrogen is primarily molecular, and the gas temperature is $\sim 30$~K; if the H$_2$Cl$^+$ originates primarily in such a region, the collisional excitation is $R_{\rm interior} = q_{\rm tot}({\rm H}_2) n_{\rm H} / 2 = 1.5 \times 10^{-6}\rm \, s^{-1}$.  Unfortunately, these estimates are quite uncertain, given the absence of reliable rate coefficients for H$_2$Cl$^+$.  In particular,  several shortcomings in the Coulomb-Born approximation have been discussed by Faure \& Tennyson (2001), and there may be significant differences in the rates of excitation of H$_2$Cl$^+$ and H$_2$O in collisions with H$_2$.

In the limit of subthermal excitation, the integrated brightness temperature of the $1_{11}- 0_{00}$ line is related to the column density of para-H$_2$Cl$^+$ by the expression

$$ 8 \pi k \int T_b dv  =  h c \lambda^2 R N({\rm p\mbox{-}H_2Cl^+}), \eqno(1)$$
where $N({\rm p\mbox{-}H_2Cl^+})$ is the beam averaged column density of para-$\rm H_2Cl^+$.
{ In deriving equation (1), we adopted the standard definition of Rayleigh-Jeans brightness temperature, $T_b = I_\nu \lambda^2/2k$, where $I_\nu$ is the specific intensity; and we assumed that the line emission is optically-thin and isotropic, with the frequency-integrated specific intensity given by $h \nu RN({\rm p\mbox{-}H_2Cl^+})/4 \pi.$}

Given an integrated { antenna} temperature of 382 mK km/s for the $1_{11}-0_{00}$ transition of H$_2^{35}$Cl$^+$, { and a main beam efficiency of 0.76 (Roelfsema et al.\ 2012)},
this expression yields an estimate of ${ 7} \times 10^{12}\, \rm cm^{-2}$ for the { beam-averaged} column density of para-H$_2^{35}$Cl$^+$, if assumed to originate in a region close to the cloud surface (where $R=R_{\rm surface}$).  Alternatively, if the para-H$_2^{35}$Cl$^+$ originates in the cloud interior (where $R=R_{\rm interior}$), the estimated column density is even higher: ${ 1.6} \times 10^{14}\, \rm cm^{-2}$.  For an assumed ortho-to-para ratio of 3 {\re for chloronium, corresponding to the equilibrium value expected at diffuse cloud temperatures,
and for} a H$_2^{35}$Cl$^+$/H$_2^{37}$Cl$^+$ ratio of 3, these values correspond to total H$_2$Cl$^+$ column densities of ${ 2.9} \times 10^{13}\, \rm cm^{-2}$ and ${ 7.0} \times 10^{14}\, \rm cm^{-2}$, respectively, for a cloud surface and cloud interior source of the observed H$_2^{35}$Cl$^+$ emission.  { These estimates both apply to the beam-averaged column density and represent two extreme cases that are expected to bracket the actual value.}

\subsection{Chloronium emission observed toward the Orion South condensation}

In the Orion South molecular condensation, the gas density is believed to be higher than in the Orion Bar PDR (and UV radiation field to be smaller.)  Mundy et al.\ (1988) estimated the virial mass of Orion S to be 80 $M_\odot$ from interferometric CS $J=2-1$  observations, and -- assuming a thickness along the line-of-sight based on the observed angular size -- obtained an estimate of $3 \times 10^6 \, \rm cm^{-3}$ for the H$_2$ density, $n({\rm H}_2) = n_{\rm H}/2$.
A better density estimate comes from fits to the $J=4-3$, $J=10-9$, $J=12-11$ and $J=16-15$
transitions of $\rm HC_3N$, observed by Bergin, Snell \& Goldsmith (1996), who mapped the entire Orion ridge in these transitions with an angular resolution of about 50$^{\prime \prime}$ .  Assuming a gas temperature of 30~K, Bergin et al.\ obtained a density estimate $n({\rm H}_2) > 1.2 \times 10^6 \, \rm cm^{-3}$ at the position of Orion S, with $n({\rm H}_2) \sim 10^6 \, \rm cm^{-3}$ at adjacent positions.
For a higher assumed gas temperature $\sim 100$~K, the density limit for Orion S would be lower by about a factor of 3.  Finally, Rodr{\'{\i}}guez-Franco et al.\ (2001) combined observations of the $N=1-0$, $N=2-1$ and $N=3-2$ lines
of CN to estimate a density $n({\rm H}_2)= 2 \times 10^5 \, \rm cm^{-3}$ for the region near Orion
S, for an assumed gas temperature of 80~K.  In our analysis of the H$_2$Cl$^+$ emission detected from Orion S, we have adopted a density estimate $n({\rm H}_2) = 3 \times 10^5 \, \rm cm^{-3}$, based on these previous values presented in the literature.

The gas temperature in Orion South has been estimated from observations of two symmetric top molecules: NH$_3$ (Batrla et al.\ 1983) and $\rm CH_3CN$ (Ziurys et al.\ 1990).  Although Batrla et al.\ mapped the (1,1) and (2,2) inversion lines of NH$_3$ with a resolution of 43$^{\prime \prime}$ , only for a few select positions were observations of the higher inversion lines made and temperatures quoted.  One such position was within 11$^{\prime \prime}$ of Orion South (their position S6).  These observations and those of Ziurys et al. are consistent with a
gas temperature of about 100 K for Orion S, a value that we adopt in deriving the H$_2$Cl$^+$ column density\footnote{We note also that
Mauersberger et al.\ (1986) detected the $(J,K) = (7,7)$ metastable
inversion line of NH$_3$, suggesting that in addition to the
100 K component, a much hotter component is also present with a
temperature $\ge$ 270~K, although based on the rotation diagram shown, less
than 10$\%$ of the NH$_3$ column density is in the high temperature
component.}.

There being no clear evidence in Orion South for a PDR {\re of the same prominence as} that in the Orion Bar, we expect collisions with H$_2$ to dominate the excitation of chloronium.  Given an integrated antenna temperature of 316~mK~km/s, and the density and temperature estimates discussed above, we derive a para-$\rm H_2^{35}Cl^+$ column density of ${ 3} \times 10^{12} \rm \, cm^{-2}$, corresponding to a total chloronium column density of ${ 1.6} \times 10^{13} \rm \, cm^{-2}$ for an assumed OPR of 3 and $\rm H_2^{35}Cl^+$/$\rm H_2^{37}Cl^+$ ratio of 3.  We estimate the beam-averaged H$_2$ column density as $4 \times 10^{23} \rm \, cm^{-2}$, based upon the 1.3~mm dust continuum map obtained by Mezger, Zylka \&
Wink (1990), which implies an $N({\rm H_2Cl^+})/N({\rm H}_2)$ abundance ratio of ${ 4} \times 10^{-11}$.  This value corresponds to a fraction $\sim { 2} \times 10^{-4}$ of the gas-phase Cl abundance {\re determined for} diffuse atomic clouds.  {\re By means of a similar analysis, we found that the upper limit on the ${\rm H_2Cl^+}$ line flux obtained toward Orion H$_2$ Peak 1 yields a limit of ${ 5} \times 10^{-10}$ on the $N({\rm H_2Cl^+})/N({\rm H}_2)$ abundance ratio at that position, a value that is a factor $\sim 10$ higher than the abundance inferred for the Orion South condensation.}

\section{Comparison with model predictions}

The past four decades have seen the publication of many theoretical studies of the chemistry of chlorine-bearing molecules in the interstellar medium (Jura 1974; Dalgarno et al. 1974; van Dishoeck \& Black 1986; Blake, Anicich \& Huntress 1985; Schilke et al.\ 1995; Federman et al.\ 1995; Amin 1996; NW09).  In diffuse molecular clouds, the chemistry is initiated by the photoionization of atomic chlorine by ultraviolet radiation in the narrow $912 - 958$ \AA\ wavelength range
$$\rm Cl + h\nu \rightarrow Cl^+ + e, \eqno(R1)$$
which is followed by the exothermic reaction
$$\rm Cl^+ + H_2 \rightarrow HCl^+ + H \eqno(R2)$$
Depending upon the molecular hydrogen fraction, $f({\rm H}_2) = 2n({\rm H}_2)/ [2n({\rm H}_2)+ n({\rm H})]$, HCl$^+$ is destroyed primarily via dissociative recombination
$$\rm HCl^+ + e \rightarrow H + Cl, \eqno(R3)$$
or by reaction with H$_2$ to form H$_2$Cl$^+$
$$\rm HCl^+ + H_2 \rightarrow H_2Cl^+ + H. \eqno(R4)$$
Chloronium ions are subsequently destroyed by dissociative recombination to form Cl or HCl:
$$\rm H_2Cl^+ + e \rightarrow Cl + {\re \rm H_2 {\rm \,\, or \,\,} 2H} \phantom{HCl + H} \eqno(R5a)$$
$$\rm H_2Cl^+ + e \rightarrow HCl + H \phantom{Cl + H_2 {\rm \,\,or\,\,} 2H}\eqno(R5b)$$

In dense molecular clouds, by contrast, where the interstellar UV radiation field is greatly attenuated, the reaction sequence is driven by cosmic ray ionization instead of photoionization.  Here, cosmic ray ionization of H$_2$ leads to the 
reactive H$_3^+$ ion (following proton transfer from H$_2^+$ to H$_2$), 
and reactions R1 and R2 are replaced by 
$$\rm Cl + H_3^+ \rightarrow HCl^+ + H_2 \eqno(R6)$$
as the primary route to HCl$^+$.

For several key reactions in the above sequence, there are no reported measurements of the rate coefficients.  In particular, the dissociative recombination (DR) rate for HCl$^+$ -- an important parameter affecting the abundances of both HCl$^+$ and H$_2$Cl$^+$ -- is unknown and urgently needed.  In their recent study of chlorine chemistry in diffuse molecular clouds, NW09 adopted a rate coefficient for this process of $2 \times 10^{-7}\, \rm cm^{3}\, s^{-1},$ a value typical of DR rate coefficients for diatomic molecular ions; however, several hydride molecular ions, including OH$^+$ and HF$^+$, have measured rate coefficients up to an order of magnitude smaller than this typical value.

\subsection{Diffuse cloud abundances toward W31C and Sgr A (+50 km/s cloud)}

To help interpret the H$_2$Cl$^+$ observations of diffuse clouds along the sight-lines to Sgr A (+50 km/s cloud) and W31C, we have used the Meudon PDR code (le Petit et al.\ 2006) to obtain predictions for the H$_2$Cl$^+$ abundance.  The approach is rather similar to that adopted by NW09, but includes an improved treatment of the depth dependence of the photochemistry and of the self-shielding of molecular hydrogen from photodissociation.  {\re In particular, the photodissociation and photoionization rates for such species as H, H$_2$, Cl, HCl, HCl$^+$ and CO have been computed by integrating the corresponding cross-sections over the radiation field intensity in an ``exact'' UV transfer approach where possible self-shielding, line overlap and continuum gas phase and dust absorption are taken into account. The contribution of the dust extinction is computed from the shape of the extinction curve whose analytic expression is given and discussed in Fitzpatrick \& Massa (1999).  The mean Galactic curve extinction has been used here as our standard assumption.  The radiation field is assumed to impinge isotropically on both sides of a slab of fixed visual extinction $A_{V,tot}$.} The reaction network is the same as that adopted by NW09. 

In Figure 4, we present the predicted fractions of gas-phase chlorine present in the HCl$^+$ and H$_2$Cl$^+$ ions, $\bar f({\rm HCl}^+)$ and $\bar f({\rm H_2Cl}^+)$, averaged along a sight-line through a diffuse molecular cloud of total visual extinction $A_{\rm V,tot}$, together with the average H$_2$ fraction, $\bar f({\rm H_2}) = 2N({\rm H}_2)/ [2N({\rm H}_2)+ N({\rm H\,I})]$.  Although the diffuse-cloud abundances of HCl$^+$ and $\rm H_2Cl^+$ show very little dependence upon the cosmic ray ionization rate, we note that the results presented in Figure 4 were computed for an assumed primary cosmic ray ionization rate of $1.8 \times 10^{-16}\, \rm s^{-1}$ per H nucleus, a value corresponding to the ``enhanced ionization rate" considered by NW09 and in agreement with recent estimates obtained from the analysis of H$_3^+$ (Indriolo et al.\ 2007) and OH$^+$ abundances (Neufeld et al.\ 2010b) in diffuse clouds.  The different panels in Figure 4 show the results for different ratios of the gas density, $n_H$, to the radiation field, $\chi_{UV}$, the latter normalized with respect to the mean radiation field given by Draine (1978).  
%The $\rm H_2Cl^+$ fractions predicted in Figure 4 are typically a factor $\sim 3$ larger than those obtained by NW09, because the improved treatment of H$_2$ self-shielding yields a larger H$_2$ fraction for a given $A_{\rm V,tot}$.  Nevertheless, 
{\re Regardless of $n_H$/$\chi_{UV}$}, the maximum value of ${\bar f}({\rm H_2Cl}^+)$ is typically obtained in clouds with $f({\rm H}_2) \sim 0.3 $.  {\re In Figure 5, we show (magenta diamonds) the ratio of the ${\rm H_2Cl^+}$ column density to that of atomic hydrogen, $N({\rm H_2Cl^+})/N({\rm H\,I})$, together with the corresponding ratio for ${\rm HCl^+}$ (blue triangles).
The maximum $N({\rm H_2Cl^+})/N({\rm H\,I})$ ratio predicted for any parameters, $1 \times 10^{-9}$, is a factor of 4 and 12 below the average values measured toward Sgr A (+50 km/s cloud) and W31C, respectively.}
A very similar discrepancy has been reported in the case of HCl$^+$, the tentative detection of which has been presented recently by De Luca et al.\ (2011).  Indeed, if we assume a  $N({\rm H_2Cl^+})/N({\rm HCl}^+)$ ratio of 1, the value implied for W31C given the De Luca et al.\ detection, the maximum predicted $N({\rm H_2Cl^+})/N({\rm H\,I})$ ratio is $5 \times 10^{-10}$, a factor of $\sim 20$ below the measured value. 

We have investigated several explanations for the {\re discrepancies} 
described above, none of which appears to be entirely satisfactory.  
First, we have examined the effect of reducing the rate coefficient assumed for DR of $\rm {\re HCl^+}$ to $2 \times 10^{-8} \,\rm cm^{3}\, s^{-1} (T/300\,{\rm K})^{-1/2}$; this value is {\re 10 times smaller} than that adopted by NW09 as the typical DR of diatomic molecular ions, but equal to the experimental value obtained by Djuri{\'c}  et al.\ (2001) for HF$^+$.  The result of this change is quite modest, however: for the cloud model that yields the maximum H$_2$Cl$^+$ abundance, with $A_{V,tot} = 0.3$ and $n_H/\chi_{UV} = 32$, the predicted HCl$^+$ and $\rm H_2Cl^+$ abundances increase by factors of 2.7 and 1.3 respectively.  Second, we investigated the effect of increasing the rate coefficient assumed for charge transfer between Cl and H$^+$.  That reaction was assumed by Blake et al.\ (1986) to have a rate coefficient of $10^{-9} \,\rm cm^{3}\, s^{-1}$ and was identified as an important route to the formation of Cl$^+$, but subsequently found to have a considerably smaller rate coefficient (by factors of $\sim 20$ and $\sim 250$ at 300~K and 100~K, {\re respectively}) in calculations performed by Pradhan \& Dalgarno (1994).  For the cloud model with $A_{V,tot} = 0.3$ and $n_H/\chi_{UV} = 32$, the predicted HCl$^+$ and $\rm H_2Cl^+$ abundances only increase by $7\%$ and 6$\%$, respectively, if a rate coefficient of $10^{-9} \,\rm cm^{3} \, s^{-1}$ is assumed in place of that computed by Pradhan \& Dalgarno (1994). 
Third, we have examined the effect of adopting an interstellar radiation field with a spectral shape corresponding to a 30,000~K blackbody in place of that assumed by Draine (1978) (but with the absolute intensity constrained to be equal at 1000 \AA\ .)  At photon energies above the ionization potential of atomic chlorine (12.97 eV), the former drops less rapidly than the latter, leading to an enhanced photoionization rate for Cl.  However, even for this harder (and arguably extreme) radiation field, {\re the resultant HCl$^+$ and $\rm H_2Cl^+$ abundances are only increased by factors of 1.4 and 1.3 respectively.}  Fourth, we have investigated the dependence of the model predictions upon the wavelength dependence of the dust opacity, assumed in our standard model to follow the average Galactic extinction law of Fitzpatrick\& Massa (1999).  We have considered a case in which the dust opacity is assumed to rise less rapidly as the wavelength decreases, following the extinction curve of HD 294264 (Fitzpatrick \& Massa 2005).  {\re Here, the peak H$_2$Cl$^+$ abundances are achieved at somewhat larger $A_{\rm V,tot}$, but the effect on the peak abundances are small: the peak predicted HCl$^+$ and $\rm H_2Cl^+$ abundances are only increased by 3$\%$ and 12$\%$ respectively.}

{\re We note two additional systematic uncertainties that might help account for the discrepancy between theory and observation.
First, the chloronium column densities derived from the absorption line observations are inversely proportional to the square of the assumed dipole moment, for which we adopted a value of 1.89~D, the result of unpublished calculations by M\"uller.  This value is already fairly large, and thus an unusually large dipole moment (larger than $\sim 4 - 5$~D) would be needed to bring the theory and observation into acceptable agreement.  Second, the chloronium column densities predicted by theory are roughly inversely proportional to the assumed DR rate for $\rm H_2Cl^+$.  The value we adopted was that obtained in recent laboratory measurements ({ Geppert \& Hamberg 2009, personal communication})
{ performed with the use of an ion storage ring}, but there is always the possibility that some degree of H$_2$Cl$^+$ rotational excitation in the laboratory environment may lead to results that differ from those applicable in the interstellar environment (where most molecules are in the ground rotational state); { in this regard, we note that Petrignani et al.\ 2011 have recently argued that studies of dissociative recombination in ion storage rings may typically involve ions with higher rotational temperatures than had previously been appreciated.}

Finally, we note that the theoretical models employed to date obtain steady-state abundances for the chlorine-bearing molecules under investigation.  Because the timescale for H$_2$ formation is relatively long --  of order 10 $(100\, {\rm cm}^{-3} / n_{\rm H})$~Myr -- time-dependent effects may be important as molecular clouds either form or disperse.  In the case of molecular cloud formation, the H$_2$ abundance lags relative to its equilibrium value, which is unfavorable for the formation of chloronium; but the case of cloud dispersal may be more favorable for the production of enhanced H$_2$Cl$^+$ abundances and is worthy of future investigation.}

\subsection{The column density in the Orion Bar PDR}

The chloronium column density predicted by NW09 (their Figure 5) for conditions appropriate to the Orion Bar PDR -- $n_H \sim 6 \times 10^4 \rm \,cm^{-2}$ and $\chi_{UV} \sim 3 \times 10^4$ ({\re Herrmann} et al.\ 1997) -- is $N({\rm H_2Cl^+}) \sim 3 \times 10^{11}\, \rm cm^{-2},$ a factor $\sim 100$ smaller than the observed value (see \S4.2, for the case of a surface source of ${\rm H_2Cl^+}$).  These two column densities are not directly comparable, however, because the theoretical prediction is of the perpendicular column density of chloronium, whereas the Orion Bar PDR is viewed nearly edge-on.  Nevertheless, the geometric enhancement in the measured column density is expected to be a factor $\sim 4$ (e.g. Neufeld et al.\ 2006), and thus the theory still appears to underpredict the observed column density by more than an order of magnitude.  Clearly, the uncertainties here are considerably larger than those involved in the comparison of theory and observation in {\it diffuse} clouds, but nevertheless it appears that the discrepancy is similar in size (and direction).

\subsection{The abundance in the Orion South condensation}

Whereas the chemistry leading to chloronium in diffuse clouds is driven by photoionization of Cl, the chemical network in dense clouds is initiated by cosmic ray ionization.  In Figure 6, we show theoretical predictions
for the ${\rm H_2Cl^+}$ abundance {\re in the shielded interior of a dense cloud}, shown here relative to that of gas-phase chlorine.  These predictions are based upon an assumed cosmic-ray ionization rate of $1.8 \times 10^{-17}\,\rm s^{-1}$ (primary ionization rate per H nucleus); this is the ``standard value'' adopted by NW09, believed to be most appropriate for dense molecular clouds.  Once again, the observed ${\rm H_2Cl^+}$ abundance of $\sim 1.5 \times 10^{-4}$ relative to the diffuse cloud gas-phase chlorine abundance (\S 4.3) is more than an order of magnitude greater than the model prediction.  The discrepancy becomes even greater if the chlorine depletion in the Orion South condensation is larger than that in the diffuse ISM, as suggested by observations of HCl by Schilke et al.\ (1997).

\section{Summary}

1.  We have detected interstellar chloronium absorption in foreground molecular clouds along the sight-lines to the bright submillimeter continuum sources Sgr A (+50 km/s cloud) and W31C.  Both para-H$_2^{35}$Cl$^+$ and para-H$_2^{37}$Cl$^+$ were detected, in absorption, in the $1_{11}-0_{00}$ ground-state transition.  For an assumed ortho-to-para ratio of 3, the observed optical depths imply that chloronium accounts for $\sim 4 - 12 \%$ of chlorine nuclei in the gas phase.  

2.  We have detected interstellar chloronium emission from two positions in the Orion Molecular Cloud I: the Orion Bar photodissociation region and the Orion South condensation.  For an assumed ortho-to-para ratio of 3, the observed emission line fluxes imply total beam-averaged column densities of $\sim { 2.9} \times 10^{13} \rm \, cm^{-2}$ and $\sim { 1.6} \times 10^{13} \rm \, cm^{-2}$, respectively, for chloronium in these two sources.

3.  The chloronium abundances inferred in this study are typically at least a factor $\sim 10$ larger than the predictions of theoretical models for the chemistry of interstellar molecules containing chlorine.  Several explanations for this discrepancy were investigated, but none has proven satisfactory, and thus the large observed abundances of chloronium remain puzzling. {\re Additional laboratory and/or theoretical investigations of the dipole moment of H$_2$Cl$^+$, its dissociative recombination rate (for rotationally-cold ions), and its rates of collisional excitation by H$_2$ and $e$ are urgently needed, as are time-dependent astrochemical models.}

\acknowledgments

HIFI has been designed and built by a consortium of institutes and university departments from across
Europe, Canada and the United States under the leadership of SRON Netherlands Institute for Space
Research, Groningen, The Netherlands and with major contributions from Germany, France and the US.
Consortium members are: Canada: CSA, U.~Waterloo; France: CESR, LAB, LERMA, IRAM; Germany:
KOSMA, MPIfR, MPS; Ireland, NUI Maynooth; Italy: ASI, IFSI-INAF, Osservatorio Astrofisico di Arcetri-
INAF; Netherlands: SRON, TUD; Poland: CAMK, CBK; Spain: Observatorio Astron\'omico Nacional (IGN),
Centro de Astrobiolog\'a (CSIC-INTA). Sweden: Chalmers University of Technology - MC2, RSS \& GARD;
Onsala Space Observatory; Swedish National Space Board, Stockholm University - Stockholm Observatory;
Switzerland: ETH Zurich, FHNW; USA: Caltech, JPL, NHSC.

Support for this work was provided by NASA through an award issued by JPL/Caltech.

{}

\begin{deluxetable}{lcccc}

\tablewidth{0pt}
\tabletypesize{\scriptsize}
\tablecaption{Summary of observations} 
\tablehead{Source & Orion Bar & Orion S & W31C & Sgr A  \\
& & & & (+50 km/s cloud)\\}
\startdata

R.A. (J2000) & \phantom{--1}5h 35m 20.60s & \phantom{--1}5h 35m 13.50s & \phantom{--}18h 10m 28.70s & \phantom{--}17h 45m 51.70s \\
Declination (J2000) & \phantom{1}--5d 25$^\prime$ 14.00$^{\prime\prime}$ & \phantom{1}--5d 24$^\prime$ 8.00$^{\prime\prime}$& 	--19d 55$^\prime$ 50.00$^{\prime\prime}$ & --28d 59$^\prime$ 09.00$^{\prime\prime}$ \\
Date & 2010-03-01 & 2010-03-01 & 2011-04-08 & 2010-09-16 \\
Total duration (s) & 4176 & 4176 & 588$^a$ & 7848$^a$ \\
Continuum $T_A$ (SSB in mK) & 66 & 800 & 382 & 115\\
Integrated line strength$^b$ (mK km/s) & 328 & 316 & &  \\
\re Line centroid $^b$ (km/s) &\re 10.5/11.3$^c$ &\re 7.8/8.5$^c$ & &  \\
\re Line FWHM$^b$ (km/s) & \re 3.4/5.1$^c$ & \re 4.0/5.7$^c$ & &  \\
\bk R.m.s noise (mK) & 4.5 & 4.5 & 9.7 & 1.4\\
Mode & Frequency scan & Frequency scan & Pointed & Pointed \\
Key program & HOP & HOP & PRISMAS & HOP \\
%\\
%Continuum $T_A$ (SSB in mK) & 66 & 382 & 115\\
%Integrated line strength$^b$ (mK km/s) & 328 & &  \\
%Line equivalent width$^b$ (km/s) & 5.0 & & \\

\enddata
\tablenotetext{a}{Divided into three observations of equal duration with different LO settings (see text)}
\tablenotetext{b}{Emission lines only}
\tablenotetext{c}{\re The first number is for a three-component Gaussian fit, with the velocity components separated in accord with the hyperfine splitting and the relative strengths appropriate to L.T.E.; the second number is for a single-component Gaussian fit, computed on the velocity scale for the strongest hyperfine component.}
\end{deluxetable}

\begin{deluxetable}{llcccccc}

\tablewidth{0pt}
\tabletypesize{\scriptsize}
\tablecaption{Chloronium column densities and abundances from absorption line observations}
\tablehead{Source & $v_{\rm LSR}$ range & $N$(p--H$_2^{35}$Cl$^+$) & $N$(p--H$_2^{37}$Cl$^+$) & $N$(H$_2$Cl$^+$) & $^{35}$Cl/$^{37}$Cl & $N_{\rm H}$ & $N$(H$_2$Cl$^+$)/$N_{\rm H}$ \\
                  & (km/s) & $\rm 10^{12}\,cm^{-2}$ & $\rm 10^{12}\,cm^{-2}$ & $\rm 10^{12}\,cm^{-2}$ &  & $\rm 10^{21}\,cm^{-2}$ & $10^{-9}$\\ }
\startdata
W31C 	& [10, 20]      & 3.9 & 1.6 & 22   & 2.5 & 2.2 & 10.0 \\
 	& [20, 30]      & 5.9 & 3.0 & 36   & 2.0 & 2.9 & 12.1 \\
 	& [30, 40]      & 8.5 & 3.2 & 47   & 2.6 & 3.5 & 13.4 \\
 	& [40, 50]      & 2.8 & 0.9 & 15   & 3.3 & 1.7 &  8.0 \\
Total: 	& [10, 50]     & 21.1 & 8.4 & 118  & 2.5 & 10 & 11.5 \\
\\

Sgr A 	& [--70, --30] & 2.3 & 0.9 & 13 & 2.5    & 2.9  &  4.4 \\
 	& [--30, +20]  & 5.5 & 1.3 & 27 & 4.2    & $> 14$ &  $< 2.0$ \\
Total: 	& [--70, +20]  & 7.8 & 2.2 & 40 & 3.5    & $> 17$ &  $< 2.4$ \\
\enddata
\tablenotetext{a}{\re The p--H$_2^{37}$Cl$^+$ absorption spectrum in this velocity range is significantly affected by an interloper absorption feature }
\end{deluxetable}

\begin{figure}
\includegraphics[width=10 cm]{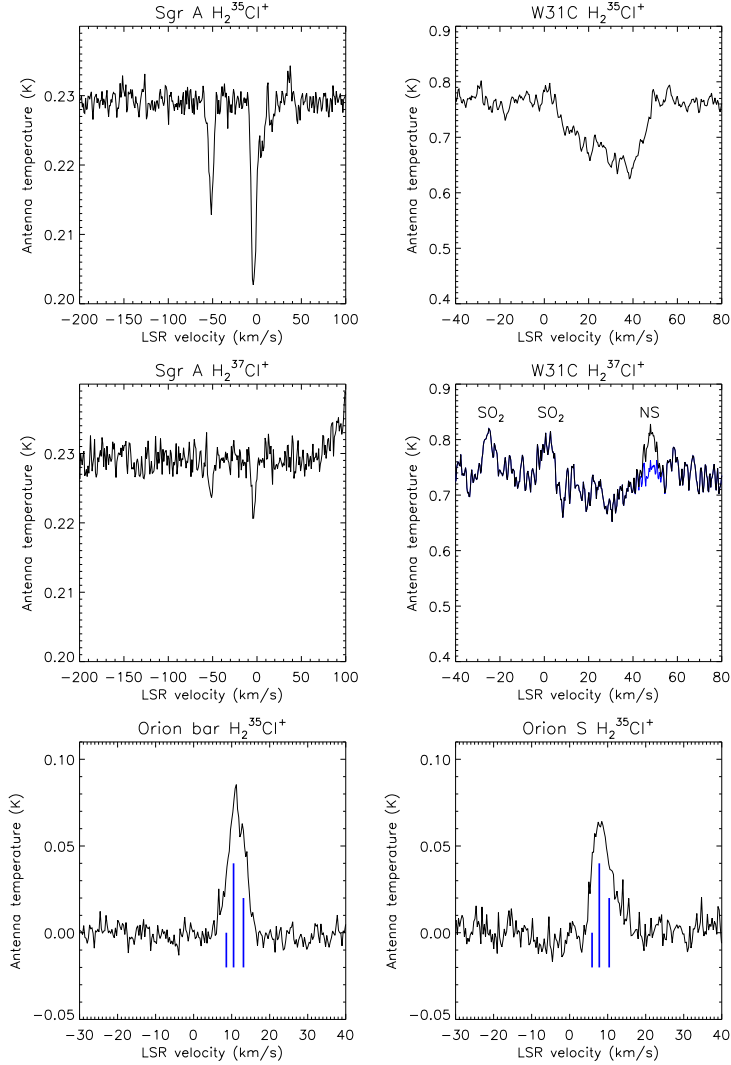}
\caption{Spectra of the para-$1_{11}-0_{00}$ transition (and, for two sources, the analogous transition of para-H$_2^{37}$Cl$^+$) obtained toward the four sources Sgr A (+50 km/s cloud), W31C, the Orion Bar, and Orion South.   Here, the intensity is plotted on the scale of antenna temperature, as a function of the LSR velocity for the strongest hyperfine components (at rest frequencies of 485.417~GHz for H$_2^{35}$Cl$^+$ and 484.232~GHz for H$_2^{37}$Cl$^+$.)  The spectra shown for Sgr A (+50 km/s cloud) and W31C are double sideband spectra, while those for the Orion Bar and Orion South are sideband deconvolved.}
\end{figure}

\begin{figure}
\includegraphics[width=10 cm]{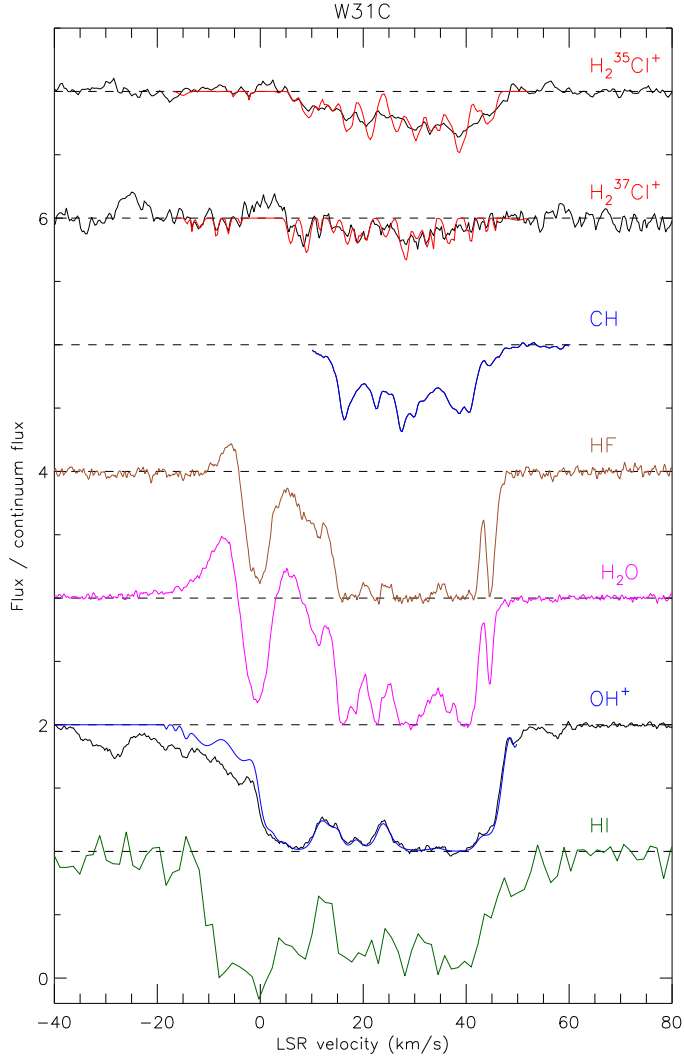}
\caption{H$_2^{35}$Cl$^+$ and H$_2^{37}$Cl$^+$ absorption spectra obtained toward W31C, normalized with respect to the continuum flux in a single sideband.  The black curves show the observed spectra, and the red curves show the result of deconvolving the hyperfine structure; these are the spectra that would have resulted if the hyperfine splitting were zero.
For comparison, the spectra of four other molecules detected with HIFI are shown: CH (the 536.761 GHz transition in green from Gerin et al. 2010b, {\re after deconvolution of the hyperfine structure}),  HF (the 1232.476 GHz transition in brown, from Neufeld et al.\ 2010a), para-H$_2$O (the 1113.343 GHz transition in magenta, from Neufeld et al.\ 2010a), and OH$^+$ (the 971.804 GHz transition, from Gerin et al.\ 2010a, both with (blue) and without (black) hyperfine deconvolution).  In addition, the H~I  21~cm spectrum is shown in dark green (from Fish et al.\ 2003).  Vertical offsets have been introduced for clarity.}
\end{figure}

\begin{figure}
\includegraphics[width=10 cm]{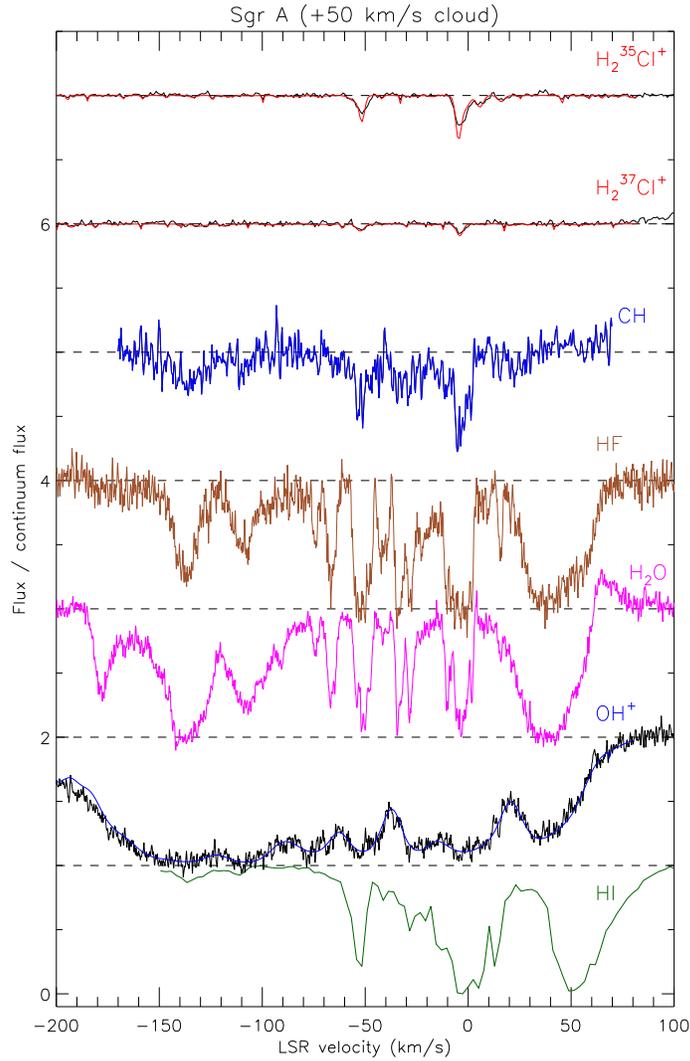}
\caption{Same as Figure 2, except for Sgr.\ A (+50 km/s cloud).  Here, the H~I  21 cm spectrum is from Lang et al.\ (2010), the OH$^+$ spectrum is from Neufeld et al.\ (2011), and the HF and H$_2$O spectra are from Sonnentrucker et al.\ (2011).  {\re The HF and H$_2$O spectra were obtained at a position 44$^{\prime\prime}$ West and 20$^{\prime\prime}$ South of that observed here.}}
\end{figure}

\begin{figure}
\includegraphics[width=15 cm]{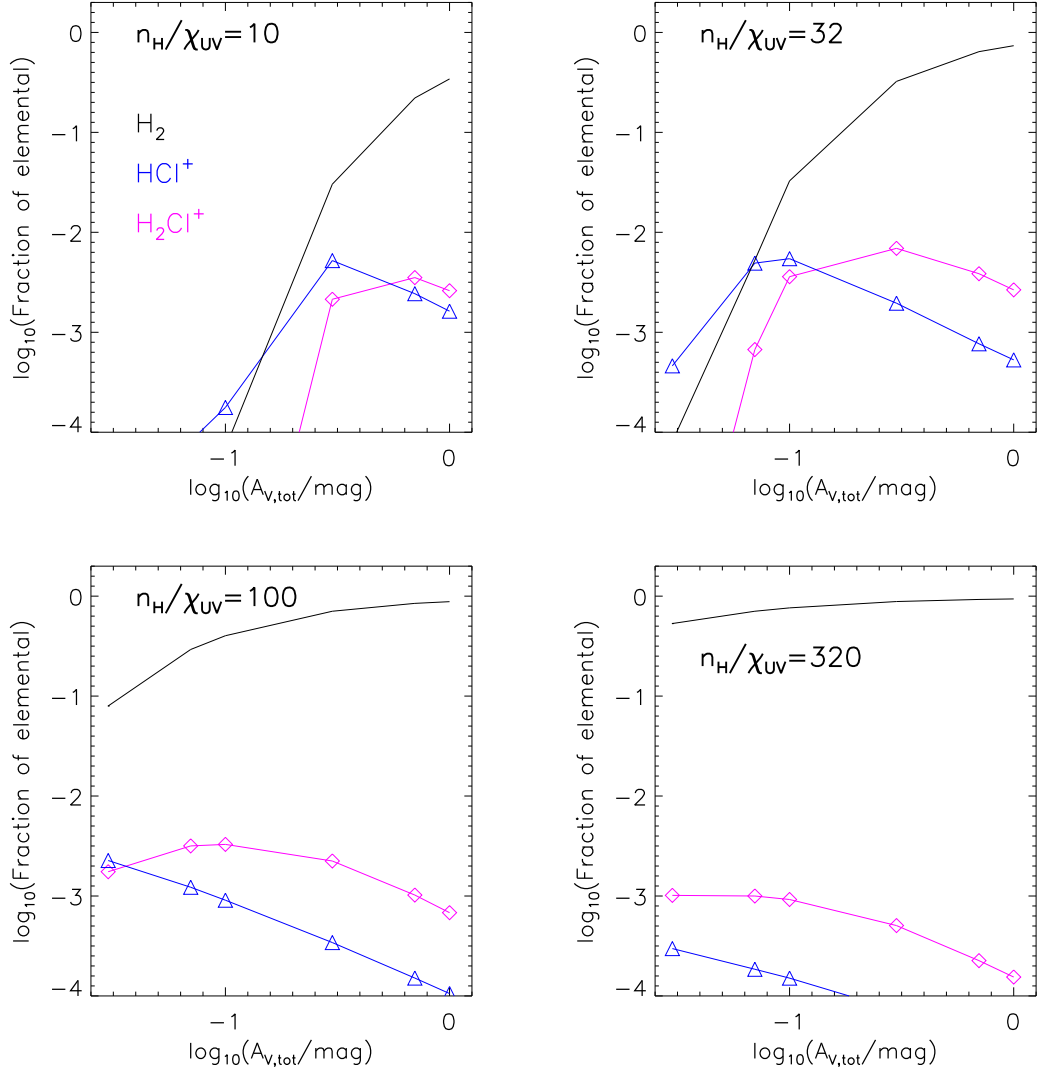}
\caption{Predicted fractions of gas-phase chlorine present in the HCl$^+$ (blue triangles) and H$_2$Cl$^+$ (magenta diamonds) ions, averaged along a sight-line through a diffuse molecular cloud of total visual extinction $A_{\rm V,tot}$, together with the average H$_2$ fraction, $\bar f({\rm H_2}) = 2N({\rm H}_2)/ [2N({\rm H}_2)+ N({\rm H\,I})]$ (black).  The different panels show the results for different ratios of the gas density, $n_{\rm H}$, to the radiation field, $\chi_{UV}$, the latter normalized with respect to the mean radiation field given by Draine (1978).}
\end{figure}

\begin{figure}
\includegraphics[width=15 cm]{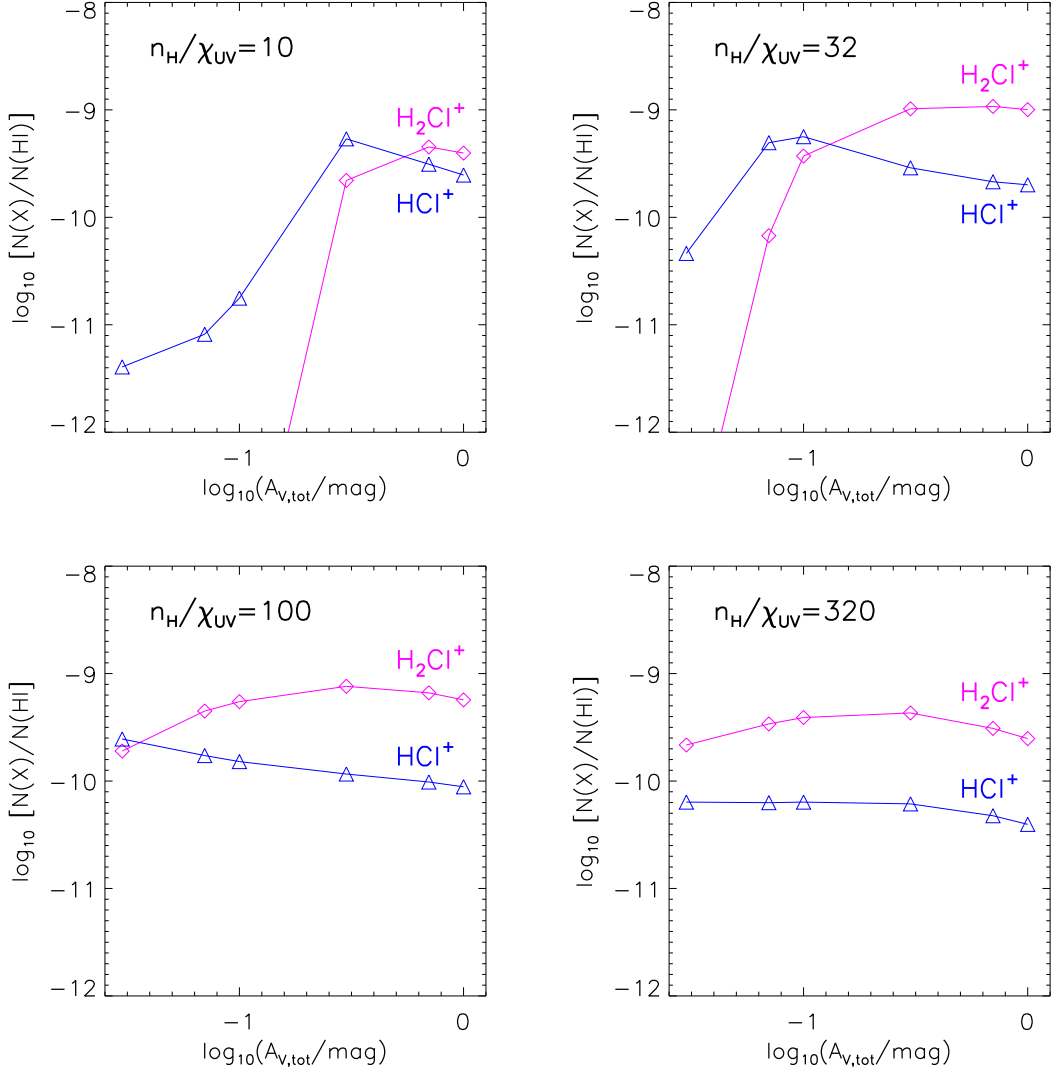}
\caption{\re Predicted $N({\rm HCl}^+)/N({\rm H\,I})$ (blue triangles) and $N({\rm H_2Cl}^+)/N({\rm H\,I})$ (magenta diamonds) ratios, averaged along a sight-line through a diffuse molecular cloud of total visual extinction $A_{\rm V,tot}$.  The different panels show the results for different ratios of the gas density, $n_{\rm H}$, to the radiation field, $\chi_{UV}$, the latter normalized with respect to the mean radiation field given by Draine (1978).}
\end{figure}

\begin{figure}
\includegraphics[width=15 cm]{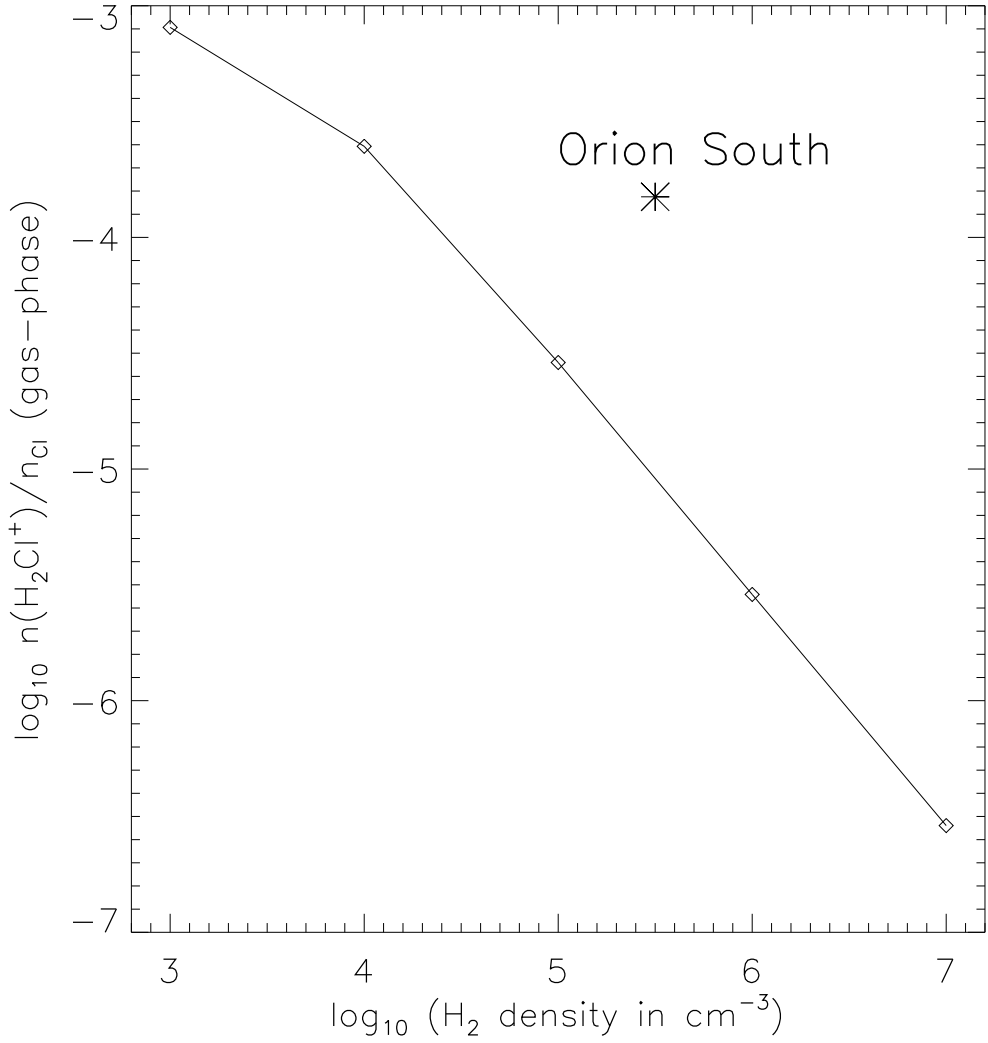}
\caption{Theoretical predictions (from NW09) for the dense cloud ${\rm H_2Cl^+}$ abundance, shown here relative to that of gas-phase chlorine, for an assumed cosmic-ray ionization rate of $1.8 \times 10^{-17}\,\rm s^{-1}$ (primary ionization rate per H nucleus)}
\end{figure}
\end{document}